\documentclass[a4paper,11pt]{article}

\usepackage{jheppub} 
\usepackage{graphicx}
\usepackage{amsmath,amssymb,amsthm,bm,bigdelim,multirow}
\usepackage{color}
\usepackage{booktabs}
\usepackage{hyperref}
\usepackage{dsfont}
\usepackage{here}
\usepackage{subcaption}

\newcommand{\kslash}{k\kern-1ex /}
\newcommand{\pslash}{p\kern-1ex /}
\newcommand{\qslash}{q\kern-1ex /}
\newcommand{\lslash}{l\kern-1ex /}
\newcommand{\sslash}{s\kern-1ex /}
\newcommand{\Dslash}{D\kern-1.2ex /}

\newcommand{\beqa}{\begin{eqnarray}}
\newcommand{\eeqa}{\end{eqnarray}}
\newcommand{\Tr}{{\rm Tr}}
\newcommand{\be}{\[}
\newcommand{\ee}{\]}
\newcommand{\bd}{\begin{description}}
\newcommand{\ed}{\end{description}}
\newcommand{\la}{\langle}
\newcommand{\ra}{\rangle}
\newcommand{\ben}{\begin{eqnarray}}
\newcommand{\een}{\end{eqnarray}}

\def\lsim{\raise0.3ex\hbox{$<$\kern-0.75em\raise-1.1ex\hbox{$\sim$}}}
\def\gsim{\raise0.3ex\hbox{$>$\kern-0.75em\raise-1.1ex\hbox{$\sim$}}}
\def\simgt{\rlap{\lower 3.5 pt\hbox{$\mathchar \sim$}}\raise 2.0pt \hbox {$>$}}
\def\simlt{\rlap{\lower 3.5 pt\hbox{$\mathchar \sim$}}\raise 2.0pt \hbox {$<$}}

\begin{document}

\title{Quantum phase transition of (1+1)-dimensional O(3) nonlinear sigma model at finite density with tensor renormalization group}

\author[a]{Xiao Luo,}
	\affiliation[a]{Graduate School of Pure and Applied Sciences, University of Tsukuba, Tsukuba, Ibaraki
    305-8571, Japan}
    	\emailAdd{luo@het.ph.tsukuba.ac.jp}

  	\author[b]{Yoshinobu Kuramashi}
  	\affiliation[b]{Center for Computational Sciences, University of Tsukuba, Tsukuba, Ibaraki
    305-8577, Japan}
  	\emailAdd{kuramasi@het.ph.tsukuba.ac.jp}

        \abstract{
          We study the quantum phase transition of the (1+1)-dimensional O(3) nonlinear sigma model at finite density using the tensor renormalization group method. This model suffers from the sign problem, which has prevented us from investigating the properties of the phase transition. We investigate the properties of the phase transition by changing the chemical potential $\mu$ at a fixed coupling of $\beta$. We determine the transition point $\mu_{\rm c}$ and the critical exponent $\nu$ from the $\mu$ dependence of the number density in the thermodynamic limit. The dynamical critical exponent $z$ is also extracted from the scaling behavior of the temporal correlation length as a function of $\mu$. 
}
\date{\today}

\preprint{UTHEP-788, UTCCS-P-156}

\maketitle

\section{Introduction}
\label{sec:intro}

The (1+1)-dimensional ((1+1)$d$) O(3) nonlinear sigma model (O(3) NLSM) is massive and shares the property of asymptotic freedom with the (3+1)$d$ non-Abelian gauge theories so that it has been used as a good testbed before exploring to investigate the properties of QCD. 
Recently we have measured the entanglement entropy and $n$th-order R\'enyi entropy for the (1+1)$d$ O(3) NLSM by directly evaluating the partition function with the tensor renormalization group (TRG) method~\footnote{In this paper, the ``TRG method" or the ``TRG approach" refers to not only the original numerical algorithm proposed by Levin and Nave~\cite{Levin:2006jai} but also its extensions~\cite{PhysRevB.86.045139,Shimizu:2014uva,PhysRevLett.115.180405,Sakai:2017jwp,PhysRevLett.118.110504,Hauru:2017tne,Adachi:2019paf,Kadoh:2019kqk,Akiyama:2020soe,PhysRevB.105.L060402,Akiyama:2022pse}.}~\cite{Luo:2023ont}. These entropies allow us to determine the central charge based on the scaling formula for the non-critical (1+1)$d$ models~\cite{Calabrese:2004eu}. This work has shown an additional ability of the TRG method to investigate the phase transition of quantum field theories from the viewpoint of the entanglement entropy.
On the other hand, in the finite density case, which may also give us useful insight into the finite density QCD, the introduction of the finite chemical potential yields a complex action problem. The previous work with the dual lattice simulation  found a quantum phase transition at the finite density regime avoiding the complex action problem and discussed the properties of the phase transition~\cite{Bruckmann:2016txt}.

In this paper we investigate the quantum phase transition of the (1+1)$d$ O(3) NLSM at finite density using the TRG method, which has been shown to be free from the sign problem or the complex action problem for various models~\cite{Shimizu:2014uva,Shimizu:2014fsa,Kawauchi:2016xng,Kawauchi:2016dcg,Yang:2015rra,Shimizu:2017onf,Takeda:2014vwa,Kadoh:2018hqq,Kadoh:2019ube,Kuramashi:2019cgs,Akiyama:2021xxr,Akiyama:2021glo,Nakayama:2021iyp,Luo:2022eje,Akiyama:2020soe,Akiyama:2020ntf,Akiyama:2023hvt}.  We performed a detailed study of the phase tranasition at a fixed value of $\beta=1.4$, which is one of the inverse coupling employed in Ref.~\cite{Luo:2023ont}. The previous study with a collective Monte Carlo algorithm obtained the correlation length $\xi_0=6.90(1)$ at zero density~\cite{Wolff:1989hv}. The corresponding mass gap is $m=0.1449(2)$ so that the cutoff effects is expected to be small at $\beta=1.4$. We determine the transition point $\mu_{\rm c}$ and the critical exponent $\nu$ from the $\mu$ dependence of number density in the thermodynamic limit. The dynamical critical exponent $z$ is also determined from the $\mu$ dependence of the temporal correlation length $\xi_t$ extracted from the ratio of the first and the second eigenvalues in the transfer matrix.
Note that $\nu=0.5$ and $z=2$ are expected from the equivalence between the (1+1)$d$ O(3) NLSM at finite density and the integer-spin Heisenberg chain with a magnetic field, whose transition when changing the magnetic field belongs to the Pokrovsky-Talapov universality class~\cite{Dzhaparidze_1978,Pokrovsky_1979,Schulz_1980,Schulz_1986,Affleck_1990}. 

This paper is organized as follows. In Sec.~\ref{sec:method}, we define the action of the (1+1)$d$ O(3) NLSM at finite density on the lattice and give the tensor network representation. 
In Sec.~\ref{sec:results} we present the numerical results for the properties of the phase transition at finite density.
Section~\ref{sec:summary} is devoted to summary and outlook.

\section{Formulation and numerical algorithm}
\label{sec:method}

Although the definition of the (1+1)$d$ O(3) NLSM at finite density and its tensor network representation are already given in the appendix of Ref.~\cite{Luo:2022eje}, we briefly give the relevant expressions for this work to make this paper self-contained.   


We consider the partition function of the O(3) NLSM on an (1+1)$d$ lattice $\Lambda_{1+1}=\{(n_1,n_2)\ \vert n_1=1,\dots,L, n_2=1,\dots,N_t\}$ whose volume is $V=L\times N_t$. The lattice spacing $a$ is set to $a=1$ unless necessary. A real three-component unit vector $\bm{s}(n)$ resides on the sites $n$ and satisfies the periodic boundary conditions $\bm{s}(n+{\hat \nu}L)=\bm{s}(n)$ ($\nu=1,2$). The lattice action $S$ is defined as
\begin{equation}
  S = -\beta \sum_{n\in\Lambda_{1+1},\nu} \sum_{\lambda,\gamma=1}^{3} s_\lambda(\Omega_n) D_{\lambda\gamma}(\mu,\hat{\nu}) s_\gamma(\Omega_{n+\hat{\nu}}),
  \label{eq:action}
\end{equation}
where the spin $s(\Omega)$ and matrix $D(\mu,\hat{\nu})$ are expressed as
\begin{align}
	& s(\Omega)= \left(\begin{array}{l}\cos\theta\\
		\sin\theta\cos\phi\\
		\sin\theta\sin\phi
	\end{array}\right), \\
	& D(\mu,\hat{\nu}) = \left(\begin{array}{rrr} 1 & ~ & ~ \\
		~ & \cosh(\delta_{2,\nu}\mu) & -i\sinh(\delta_{2,\nu}\mu) \\
		~ & i\sinh(\delta_{2,\nu}\mu )& \cosh(\delta_{2,\nu}\mu)
	\end{array}\right) 
\end{align}
with 
\begin{equation}
	\label{U:representation}
	\begin{array}{c}
		 \Omega=(\theta,\phi) \quad, ~\theta \in (0,\pi],~\phi \in (0,2\pi]. 
	\end{array}
\end{equation}
Note that we introduce the chemical potential to the rotation between the second and third components.




The partition function and its measure are written as
\begin{align}
	\label{eq:partitionfunction2}
	Z &= \int {\cal D}\Omega \prod_{n,\nu} e^{\beta  \sum_{\lambda,\gamma=1}^{3}s_\lambda(\Omega_n) D_{\lambda\gamma}(\mu,\hat{\nu}) s_\gamma(\Omega_{n+\hat{\nu}})}, \\
	{\cal D}\Omega &= \prod_{p=1}^{V} \frac{1}{4\pi} \sin(\theta_p) d\theta_p d\phi_p~.
\end{align}
We discretize the integration (\ref{eq:partitionfunction2}) with the Gauss-Legendre quadrature~\cite{Kuramashi:2019cgs,Akiyama:2020ntf} after changing the integration variables:
\ben
-1 \le \alpha&=&\frac{1}{\pi}\left(2\theta-\pi \right)\le 1, \\
-1 \le \beta&=&\frac{1}{\pi}\left(\phi-\pi \right)\le 1. 
\een
We obtain
\begin{equation}
	Z = \sum_{ \{\Omega_1\},\cdots,\{\Omega_V\}} \left( \prod_{n=1}^{V} \frac{\pi}{8}  \sin(\theta(\alpha_{a_n})) w_{a_n} w_{b_n} \right) \prod_{\nu} M_{\Omega_n,\Omega_{n+\hat{\nu}}}
\end{equation}
with $\Omega_n=(\theta(\alpha_{a_n}),\phi(\beta_{b_n}))\equiv (a_n,b_n)$, where $\alpha_{a_n}$ and $\beta_{b_n}$ are $a$- and $b$-th roots of the $K$-th Legendre polynomial $P_{K}(s)$ on the site $n$, respectively. $\sum_{ \{\Omega_n\}}$ denotes $\sum_{a_n=1}^{K}\sum_{b_n=1}^{K}$.
$M$ is a 4-legs tensor defined by
\begin{equation}
	M_{a_n,b_n,a_{n+\hat{\nu}}, b_{n+\hat{\nu}}}  =\exp\left\{ \beta \sum_{\lambda,\gamma=1}^{3}s_\lambda(a_n,b_n) D_{\lambda\gamma}(\mu,\hat{\nu}) s_\gamma (a_{n+\hat{\nu}}, b_{n+\hat{\nu}}) \right\}~.
\end{equation} 
The weight factor $w$ of the Gauss-Legendre quadrature is defined as
\begin{equation}
	w_{a_n} = \frac{2(1-{\alpha_{a_n}}^2)}{K^2P^2_{K-1}({\alpha_{a_n}})},\quad
	w_{b_n} = \frac{2(1-{\beta_{b_n}}^2)}{K^2P^2_{K-1}({\beta_{b_n}})}.
\end{equation}
After performing the singular value decomposition (SVD) on $M$:
\begin{equation}
	M_{a_n,b_n,a_{n+\hat{\nu}}, b_{n+\hat{\nu}}} \simeq \sum_{i_n=1}^{D_\text{cut}} U_{a_n,b_n, i_n} (\nu) \sigma_{i_n}(\nu) V^\dagger_{i_n,a_{n+\hat{\nu}}, b_{n+\hat{\nu}}} (\nu),
\end{equation}
where $U$ and $V$ denotes unitary matrices and $\sigma$ is a diagonal matrix with the singular values of $M$ in the descending order.
We can obtain the tensor network representation of the O(3) NLSM on the site $n\in\Lambda_{1+1}$
\begin{align}
	T_{x_n, x'_n, y_n, y'_n} &= \frac{\pi}{8} \sqrt{\sigma_{x_n}(1) \sigma_{x'_n}(1) \sigma_{y_n}(2) \sigma_{y'_n}(2) } \sum_{a_n, b_n} w_{a_n} w_{b_n} \sin(\theta_{a_n}) \nonumber \\
	&\quad \times V^\dagger_{x_n,a_n,b_n} (1) U_{a_n,b_n, x'_n} (1) V^\dagger_{y_n,a_n,b_n} (2) U_{a_n,b_n, y'_n} (2) , 
\end{align} 
where the bond dimension of tensor $T$ is given by  $D_{\text{cut}}$, which controls the numerical precision in the TRG method. The tensor network representation of partition function is given by
\begin{equation}
  Z \simeq \sum_{x_0 x'_0 y_0 y'_0 \cdots} \prod_{n \in \Lambda_{1+1}} T_{x_n x'_n y_n y'_n} = \Tr \left[T \cdots T\right]~.
  \label{eq:Z_TN}
\end{equation}
In order to evaluate $Z$ we employ the higher order tensor renormalization group (HOTRG) algorithm~\cite{PhysRevB.86.045139}.

\begin{figure}[H]
	\centering
	\includegraphics[angle=0, width=80mm]{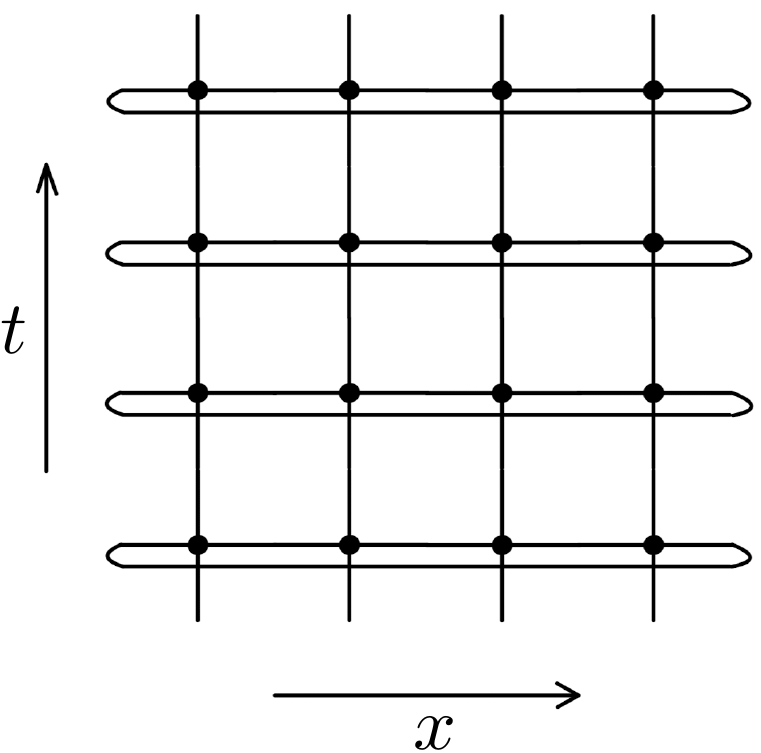}
\caption{Density matrix on a $4\times 4$ lattice. Periodic boundary condition is taken in the spatial direction.}
        \label{fig:den_cal}
 \end{figure}

The density matrix is defined by $\rho=\Tr_{x} [T\cdots T]$, where $\Tr_x$ denotes the trace in terms of the spatial direction of the legs of tensors. Figure~\ref{fig:den_cal} illustrates the density matrix in the case of a $4\times 4$ lattice.  
The temporal correlation length is obtained by
\ben
\xi_t = {N_t \over \ln\left(\frac{\lambda_0}{\lambda_1}\right)}, 
\een
where $\lambda_0$ and $\lambda_1$ is the largest and the second largest eigenvalues of the density matrix $\rho_{yy^\prime}=\Tr_x T^*_{xxyy^\prime}$ with $T^*$ the reduced single tensor obtained by HOTRG.

\section{Numerical results} 
\label{sec:results}

We calculate the partition function $Z$ at the fixed value of $\beta=1.4$ using the HOTRG algorithm with the bond dimension $D_{\rm cut}=125$, 130 and  135.  The free energy density is obtained by
\be
f=-\frac{1}{L N_t}\ln Z.
\ee
The parameter space of the three-component unit vector is discretized with $K=100$, which gives much smaller uncertainties than $D_{\rm cut}$.
We evaluate the number density with the numerical differentiation of $f$:
\be
\langle n\rangle=\frac{\partial}{\partial \mu}f\approx \frac{-1}{L N_t}\frac{\ln Z(\mu+\Delta\mu)-\ln Z(\mu-\Delta\mu)}{2\Delta\mu}.
\ee

In order to discuss the phase transition we introduce $\delta\equiv \vert \mu-\mu_{\rm c}\vert $ to measure the distance from the transition point $\mu_{\rm c}$. The correlation length in the space direction, which is denoted by $\xi$, should diverge as $\delta^{-\nu}$ with the critical exponent $\nu$ at the criticality of the second order phase transition. If the system has the symmetry between the spatial and temporal directions, the spatial correlation length should be the same as the temporal one $\xi_t$. The current model defined in Eq.~(\ref{eq:action}), however, breaks the symmetry due to the introduction of the chemical potential so that $\xi_t$ should be deviated from $\xi$. The relation between two correlation lengths are given by $\xi_t\sim \xi^z\sim \delta^{-z\nu}$ with $z$ the dynamical critical exponent.

\begin{figure}[H]
	\centering
	\includegraphics[angle=0, width=100mm]{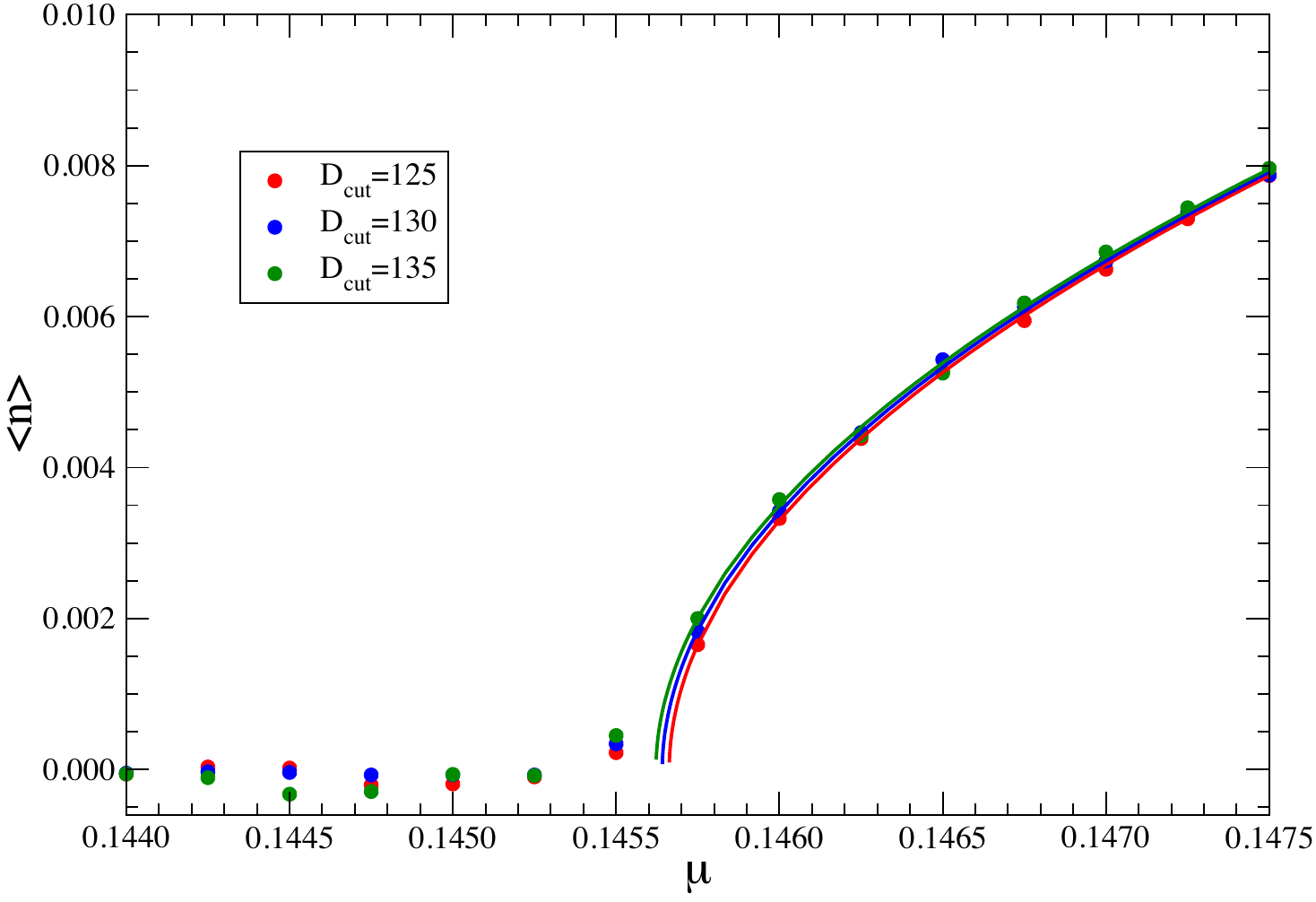}
	\caption{$\mu$ dependence of number density $\langle n\rangle$ at $\beta=1.4$ on a $2^{25}\times 2^{25}$ lattice. The bond dimension is $D_{\rm cut}\in[125,135]$. The solid curves represent the fit results at $D_{\rm cut}=125$(red), 130(blue) and 135(green).}
  	\label{fig:n}
\end{figure}

Figure~\ref{fig:n} shows the $\mu$ dependence of the number density $\langle n\rangle$ in the vicinity of the transition point at $\beta=1.4$ on a $V=L\times N_t=2^{25}\times 2^{25}=33554432\times 33554432$ lattice with $D_{\rm cut}\in [125,135]$. The temperature and spatial extension normalized by the mass gap are given by $T/m=2.1\times 10^{-7}$ and $Lm=4.9\times 10^{6}$, which is large enough to be regarded as the thermodynamic limit at zero temperature. We observe the Silver Blaze phenomenon in Fig.~\ref{fig:n}: the vanishing number density $\la n\ra$ up to $\mu\approx 0.1455$ and sudden increase of $\la n\ra$ beyond $\mu_{\rm c}$. Although the results with $D_{\rm cut}=125$,  130 and 135 are almost degenerate, we find slight $D_{\rm cut}$ dependence. We apply the global fit to the data at $D_{\rm cut}\in [125,135]$ employing the function of $\langle n\rangle (\mu,D_{\rm cut})=A_n\cdot \left\{\mu-(\mu_{\rm c}+B_n/D_{\rm cut})\right\}^\nu$ with $A_n$, $\mu_{\rm c}$, $B_n$ and $\nu$ the fit parameters. The fit range is chosen to be $0.14575\le \mu\le 0.14700$.  The solid curves show the fit results with $A_n=0.20(2)$, $\mu_{\rm c}=0.14512(11)$, $B_n=0.068(12)$ and $\nu=0.512(15)$ at $D_{\rm cut}=125$, 130 and 135. Note that the value of $\mu_{\rm c}$ is consistent with the mass gap $m=1/\xi_0=1/6.90(1)=0.1449(2)$ at $\mu=0$ obtained by a high precision Monte Carlo result with a collective algorithm~\cite{Wolff:1989hv}.

\begin{figure}[H]
	\centering
	\includegraphics[angle=0, width=100mm]{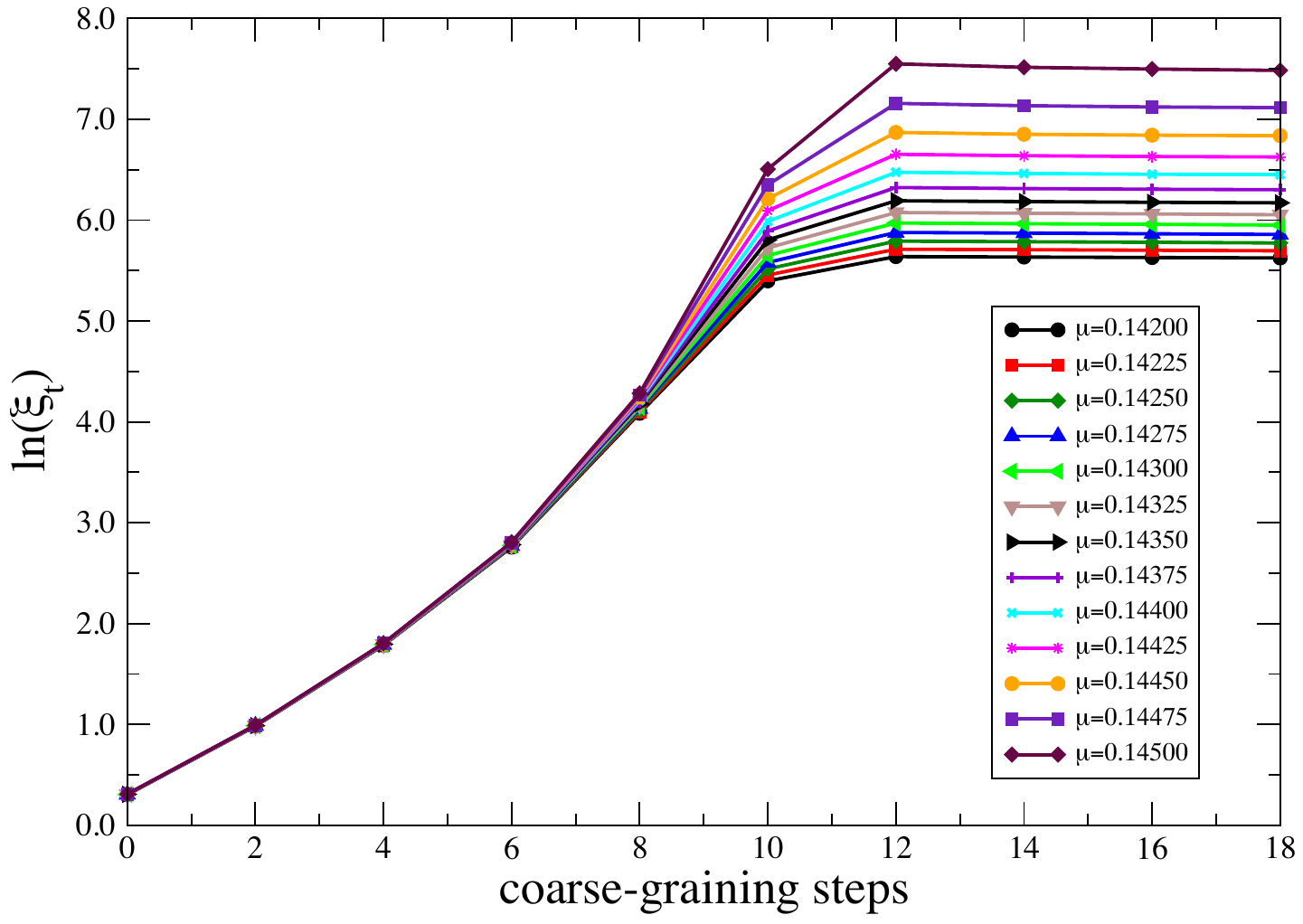}
	\caption{Temporal correlation length $\ln(\xi_t)$ at $\beta=1.4$ as a function of the coarse-graining steps. The bond dimension is $D_{\rm cut}=135$.}
  	\label{fig:xi_t_rg}
\end{figure}

\begin{figure}[H]
	\centering
	\includegraphics[angle=0, width=100mm]{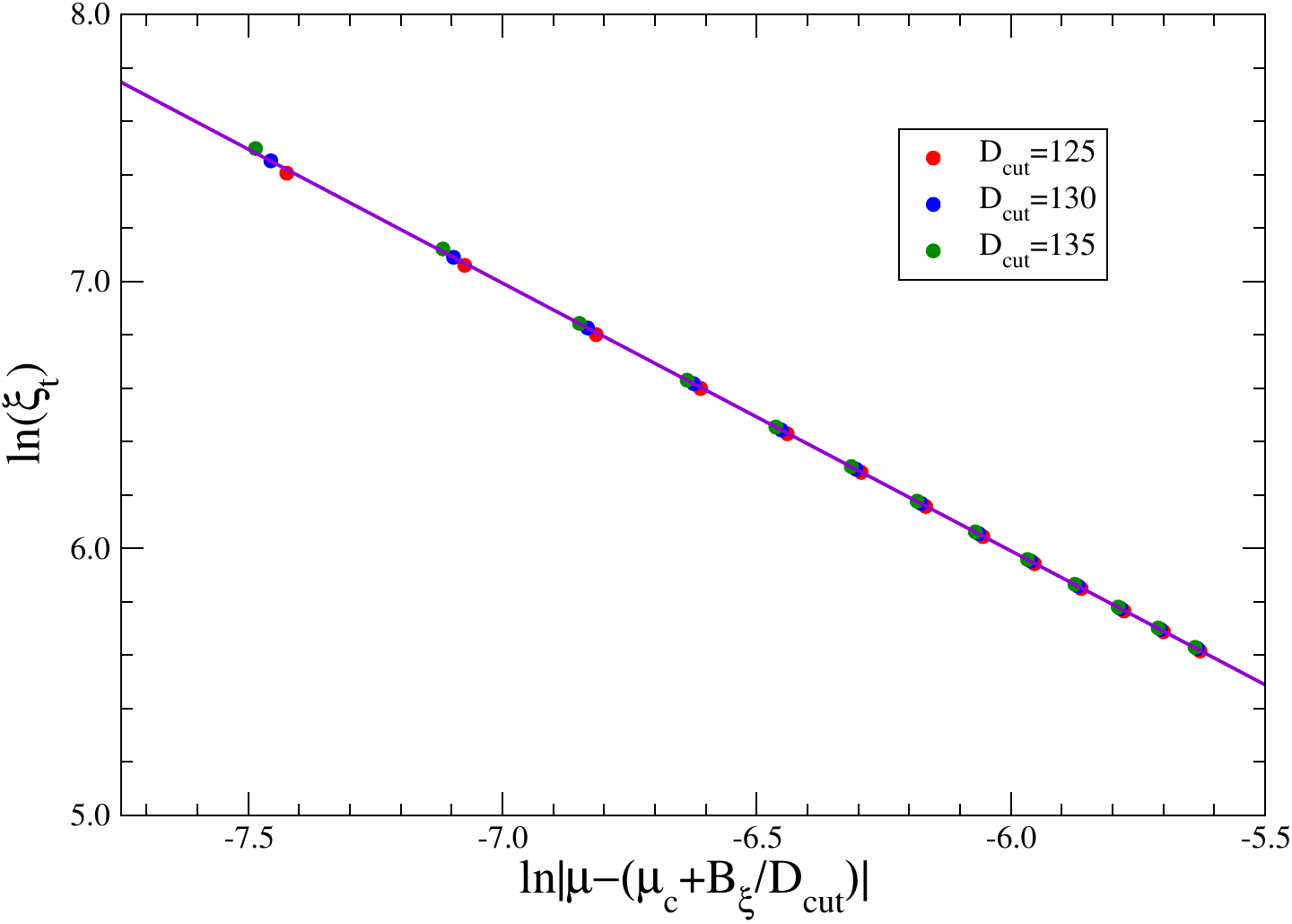}
        \caption{Temporal correlation length $\ln(\xi_t)$ at $\beta=1.4$ as a function of $\ln\vert \mu-(\mu_{\rm c}+B_\xi/D_{\rm cut})\vert$. The bond dimension is $D_{\rm cut}\in[125,135]$. The solid curve represents the fit result choosing $D_{\rm cut}=\infty$.}
  	\label{fig:xi_t_mu}
\end{figure}

The temporal correlation length $\xi_t$ is calculated from the largest and second largest eigenvalues $\lambda_0$ and $\lambda_1$, respectively, of the density matrix as already explained at the end of Sec.~\ref{sec:method}. Figure~\ref{fig:xi_t_rg} plots $\ln(\xi_t)$ as a function of the number of the coarse-graining steps at $\beta=1.4$ with $D_{\rm cut}=135$. The temporal correlation length shows plateau behavior once the spatial lattice size becomes larger than $\xi$ at the sufficiently low temperature. The plateau values increases as $\mu$ approaches the critical value of $\mu_{\rm c}=0.14512$. We make a fit of the temporal correlation length, which is measured with the density matrix of 16-th coarse-graining step, and employ the fit form of $\ln \xi_t(\mu,D_{\rm cut})=A_\xi+\alpha \ln|\mu-(\mu_{\rm c}+B_\xi/D_{\rm cut})|$ with $\mu_{\rm c}=0.14512$. The fit parameters $A_\xi$, $B_\xi$ and $\alpha$ are determined to be $A_\xi=-0.030(29)$, $B_\xi=0.0599(9)$ and $\alpha=1.003(5)$. In Fig.~\ref{fig:xi_t_mu} we plot $\ln \xi_t(\mu,D_{\rm cut})$ as a function of  $\ln|\mu-(\mu_{\rm c}+B_\xi/D_{\rm cut})|$ with $D_{\rm cut}\in [125,135]$. The solid curve represents the fit result with the choice of $D_{\rm cut}=\infty$, which shows fairly linear behavior. Using the relation of $\alpha=z\nu$ with $\nu=0.512(15)$ we obtain the dynamical critical exponent $z=1.96(6)$.

The critical exponents $\nu$ and $z$ were previously studied using the dual lattice simulation~\cite{Bruckmann:2016txt}. Although the values of $\nu$ and $z$ were not directly determined, the authors observed that the assumption of $\nu=0.5$ and $z=2$ reasonably explains the results for the scaling properties of the spin stiffness. On the other hand, our results of $\nu=0.512(15)$ and $z=1.96(6)$ directly show the consistency with the theoretical expectation of $\nu=0.5$ and $z=2$.

\section{Summary and outlook} 
\label{sec:summary}

We have investigated the properties of the quantum phase transition of the (1+1)$d$ O(3) NLSM at finite density with the TRG method. The critical chemical potential $\mu_{\rm c}=0.14512(11)$ at $\beta=1.4$ in the limit of $D_{\rm cut}\rightarrow \infty$ is consistent with the mass gap $m=1/\xi_0=1/6.90(1)=0.1449(2)$ at $\mu=0$ obtained in the previous work with a collective Monte Carlo algorithm~\cite{Wolff:1989hv}. We have also determined the critical exponent $\nu$ and the dynamical one $z$ for the transition. Our results of $\nu=0.512(15)$ and $z=1.96(6)$ are consistent with the theoretical expectation of $\nu=0.5$ and $z=2$.
This is the first successful calculation of the dynamical critical exponent with the TRG method. We are now ready to investigate the properties of the quantum phase transitions of other finite density systems.

\begin{acknowledgments}
  Numerical calculation for the present work was carried out with the supercomputers Cygnus and Pegasus under the Multidisciplinary Cooperative Research Program of Center for Computational Sciences, University of Tsukuba. We also used the supercomputer Fugaku provided by RIKEN through the HPCI System Research Project (Project ID: hp220203, hp230247).
This work is supported in part by Grants-in-Aid for Scientific Research from the Ministry of Education, Culture, Sports, Science and Technology (MEXT) (Nos. 24H00214, 24H00940).

\end{acknowledgments}



\bibliographystyle{JHEP}
\bibliography{bib/formulation,bib/algorithm,bib/discrete,bib/grassmann,bib/continuous,bib/gauge,bib/review,bib/for_this_paper}

\end{document}